\documentclass[aps,prb,twocolumn,showpacs,10pt,superscriptaddress,floatfix,longbibliography]{revtex4-1}

\usepackage{mathptmx}
\usepackage{graphicx}
\usepackage{subfigure}
\usepackage{calrsfs}
\DeclareMathAlphabet{\pazocal}{OMS}{zplm}{m}{n}
\usepackage{amsmath}
\usepackage{amsfonts}
\usepackage{amssymb}
\usepackage{mathrsfs}
\usepackage{bm}
\usepackage{subfigure}
\usepackage{xcolor}
\usepackage[colorlinks=true,linkcolor=blue,anchorcolor=red,citecolor=blue, urlcolor=blue]{hyperref}\usepackage{ulem}

\newlength{\seplinewidth}
\newlength{\seplinesep}
\colorlet{sepline}{orange}

\begin{document}

\title{Topological superconducting domain walls in magnetic Weyl semimetals}

\author{Zhao Huang}
\affiliation{Theoretical Division, Los Alamos National Laboratory, Los Alamos, New Mexico 87545, USA}

\author{Christopher Lane}
\affiliation{Theoretical Division, Los Alamos National Laboratory, Los Alamos, New Mexico 87545, USA}
\affiliation{Center for Integrated Nanotechnology, Los Alamos National Laboratory, Los Alamos, New Mexico 87545, USA}

\author{Dmitry Yarotski}
\affiliation{Center for Integrated Nanotechnology, Los Alamos National Laboratory, Los Alamos, New Mexico 87545, USA}

\author{A. J. Taylor}
\affiliation{Center for Integrated Nanotechnology, Los Alamos National Laboratory, Los Alamos, New Mexico 87545, USA}

\author{Jian-Xin Zhu}
\email{jxzhu@lanl.gov}
\affiliation{Theoretical Division, Los Alamos National Laboratory, Los Alamos, New Mexico 87545, USA}
\affiliation{Center for Integrated Nanotechnology, Los Alamos National Laboratory, Los Alamos, New Mexico 87545, USA}

\begin{abstract}
Recent experimental breakthrough in magnetic Weyl semimetals have inspired exploration on the novel effects of various magnetic structures in these materials. Here we focus on a domain wall structure which connects two uniform domains with different magnetization directions. We study the topological superconducting state in presence of an $s$-wave superconducting pairing potential. By tuning the chemical potential, we can reach a topological state, where a chiral Majorana mode or zero-energy Majorana bound state is localized at the edges of the domain walls. This property allows a convenient braiding operation of Majorana modes by controlling the dynamics of domain walls.
\end{abstract}
\date{\today}
\maketitle


{\it Introduction} -- 
The last decade has witnessed spectacular advancements in the understanding, prediction, and synthesis of new topological materials with diverse exotic emergent phenomena originating from the nontrivial band topology. One of the most interesting classes is the Weyl semimetal where the low-energy electronic dispersion closely follows the three dimensional Weyl Hamiltonian~\cite{RevModPhys18}. The salient properties originate from the nontrivial Berry curvature exhibited around the nodal points in the three-dimensional band structure. Moreover, by integrating the Berry curvature over a closed manifold around a given node yields its topological charge and chirality. Recently, magnetic Weyl semimetals have gained great interest since the broken time-reversal symmetry prevents the Weyl nodes with opposite chirality from overlapping and annihilating one another \cite{Wan11,Xu11,Wang16,Kuroda17,Liu19,Belopolski19,Morali19,Puphal20,Gao21}. Thus, this new class of materials provide a new platform to study the interplay between nontrivial band topology, magnetism, and correlations, thereby opening up new routes to novel quantum phenomena.

Magnetic Weyl semimetals are expected to host topological superconductivity when put in proximity to a trivial superconductor~\cite{Meng12, Yang14, Burkov15, Chen16,Alidoust17,Li18, Nakai20}. This delicate state is predicted to support chiral Majorana modes or localized Majorana bound states at  edges, endings or defects~\cite{RevModPhys11,Alicea_2012,Sato_2017}. These zero-energy Majorana modes mimic the Majorana fermion and can follow non-Abelian statistics, making them an ideal building block for decoherence-free quantum computers~\cite{Kitaev_2001,Milestone16}. However, despite recent advancements \cite{Fu08,Sato09,Lutchyn10,PhysRevB.92.115119,PhysRevX.7.021032,Wang18,Deng1557,Ren19,Deng12,PhysRevX.7.021011,Feng20}, a framework for effective and robust control of the individual Majorana modes is still elusive. Furthermore, with the ever growing number of newly synthesized magnetic Weyl semimetals \cite{Kuroda17,Liu19,Belopolski19,Morali19,Puphal20,Gao21}, engineering Majorana modes in these promising compounds has become a pressing and urgent issue.

Controlled magnetic switching and domain wall motion have formed a basis for magnetic random access memory \cite{Chappert07} and magnetic nanowire device concepts \cite{Allwood1688, Parkin190}, which lie in the core of spintronics. Large metastable magnetic domains have also been observed in a noncentrosymmetric feerromagnetic Weyl semimetal candidate CeAlSi \cite{Xu21}. Therefore, the existence and control of these magnetic domain walls in magnetic Weyl semimetals provides a hope to realize new electronic states.

In this Letter, we propose that a magnetic Weyl semimetal with magnetic domain walls in proximity to a superconducting pairing potential can host chiral Majorana edge modes, or zero-energy Majorana bound states, at the edge of the domain walls. The appearance of these modes follows from the highly anisotropic dependence of the nontrivial band topology on the direction of the magnetization. Specifically, we show that the magnetization orientation induced area difference of nontrivial topology in the phase diagram can be utilized to enable topologically trivial superconductivity away from the domain wall, whereas a topologically nontrivial superconductivity on the domain wall, thus realizing localized topological edge states on the domain walls. Lastly, we demonstrate that the braiding of Majorana modes can be straightforwardly achieved by tuning the magnetization of the domains.

{\it Method} -- The effective Hamiltonian of a magnetic Weyl semimetal may be written for a cubic lattice as
\begin{eqnarray}\label{H0}
H_0 &&=\sum_j -\mu c_j^{\dagger}c_j+\left[ -t_xc_j^\dagger \sigma_x c_{j+\hat{x}}-t(c_j\sigma_x c_{j+\hat{y}}+c_j\sigma_x c_{j+\hat{z}})\right. \nonumber
 \\ 
&&\left.-it'( c_j^\dagger \sigma_y c_{j+\hat{y}}+c_j^\dagger \sigma_z c_{j+\hat{z}})+H.c.\right] \nonumber \\
&&+m_xc_j^\dagger \sigma_x c_j+m_zc_j\sigma_z c_j, 
\end{eqnarray}
where $\mu$ is the chemical potential, $t_x$, $t$, and $t'$ are the coefficients of spin-orbit coupling, $m_x$ and $m_z$ are the projected Zeeman field strengths on the $x$ and $z$ axes, respectively. By inspection, time-reversal and inversion symmetry are both broken in this Hamiltonian. Upon Fourier transforming, the energy eigenvalues are $E=-\mu\pm\sqrt{h_x^2+h_y^2+(h_z+m_z)^2}$ with $h_x=m_x-2t_x\cos k_x -2t \cos k_y-2t\cos k_z$, $h_y=2t'\sin k_y$ and $h_z=2t'\sin k_z$, where the position of the Weyl nodes satisfies $h_x=h_y=h_z+m_z=0$.

To include the presence of superconductivity, we add the pairing potential in the mean-field,
\begin{eqnarray}\label{Hsc}
H_{sc} &&=\sum_j [\Delta c_{j\uparrow}^\dagger c_{j\downarrow}^\dagger+\Delta^* c_{j\downarrow}c_{j\uparrow}]\;,
\end{eqnarray}
where $\Delta$ is a momentum independent uniform s-wave gap function. Here we do not restrict the source of superconductivity. It can be either intrinsic or induced by proximity.

\begin{figure}[t]
\begin{center}
\includegraphics[clip = true, width =\columnwidth]{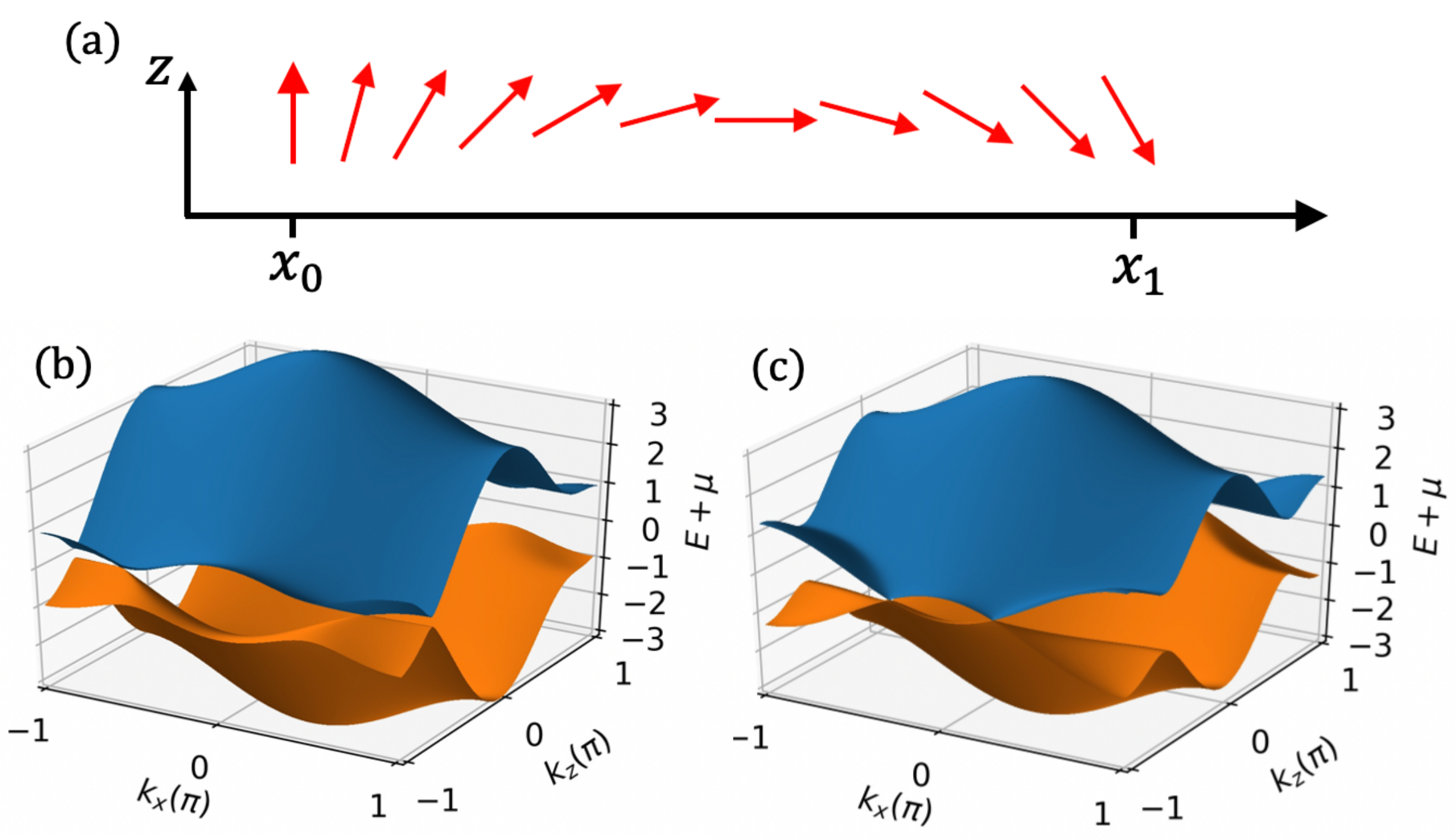}
\caption{(Color) Schematic of magnetization rotation on a domain wall (a).  The electronic band structure of a generic magnetic Weyl semimetal at $k_y=0$ for (b) Weyl phase I [$\hat{\mathbf{m}}=m_z \hat{z}$] and (c) Weyl phase II [$\hat{\mathbf{m}}=m_x \hat{x}$]. }
\label{fig1}
\end{center}
\end{figure}

To model a domain wall in this system, we allow $m_x$ and $m_z$ to be spatially dependent over the whole $x-z$ plane. Without loss of generality, we take $m_z=m \cos[Q(x-x_0)]$ and $m_x=m \sin[Q(x-x_0)]$, where the magnetization rotates through an angle of $Q(x-x_0)$ between points $x_0$ and $x$, as shown in Fig. \ref{fig1}(a).  Since translational invariance is broken along the $x$-axis, the domain wall system is governed by the two-dimensional topological classification rules. Consequently, the topological charge is obtained from the Chern number of the $k_y-k_z$ plane in the momentum space. Furthermore, the eigen-energy and wavefunction of the edge states are obtained within a thin-film geometry, which leads to a quasi-one-dimensional domain wall.

{\it Results} -- Figure~\ref{fig1}(b) and \ref{fig1}(c) show the band structure at $k_y=0$ for the magnetization along the $z$ and $x$ axes, respectively, with no superconductivity. The hopping parameters $t_x=0.4t$, $t'=0.4t$ were used. For the Zeeman field along the $z-$axis $[m_z=0.5t]$, four Weyl nodes are present at $\mathbf{k}=(\pm0.68\pi, 0, -0.79\pi)$ and $(\pm0.32\pi, \pi, -0.21\pi)$ in the Brillouin zone. For the magnetization along the $x$-axis $[m_x=0.5t]$, the four Weyl nodes are located at $(\pm0.29\pi, 0,\pi)$ and $(\pm 0.29\pi,\pi,0)$. These two cases are called ``Weyl phase I" and ``Weyl phase II" for convenience.  

\begin{figure}[t]
\begin{center}
\includegraphics[clip = true, width =\columnwidth]{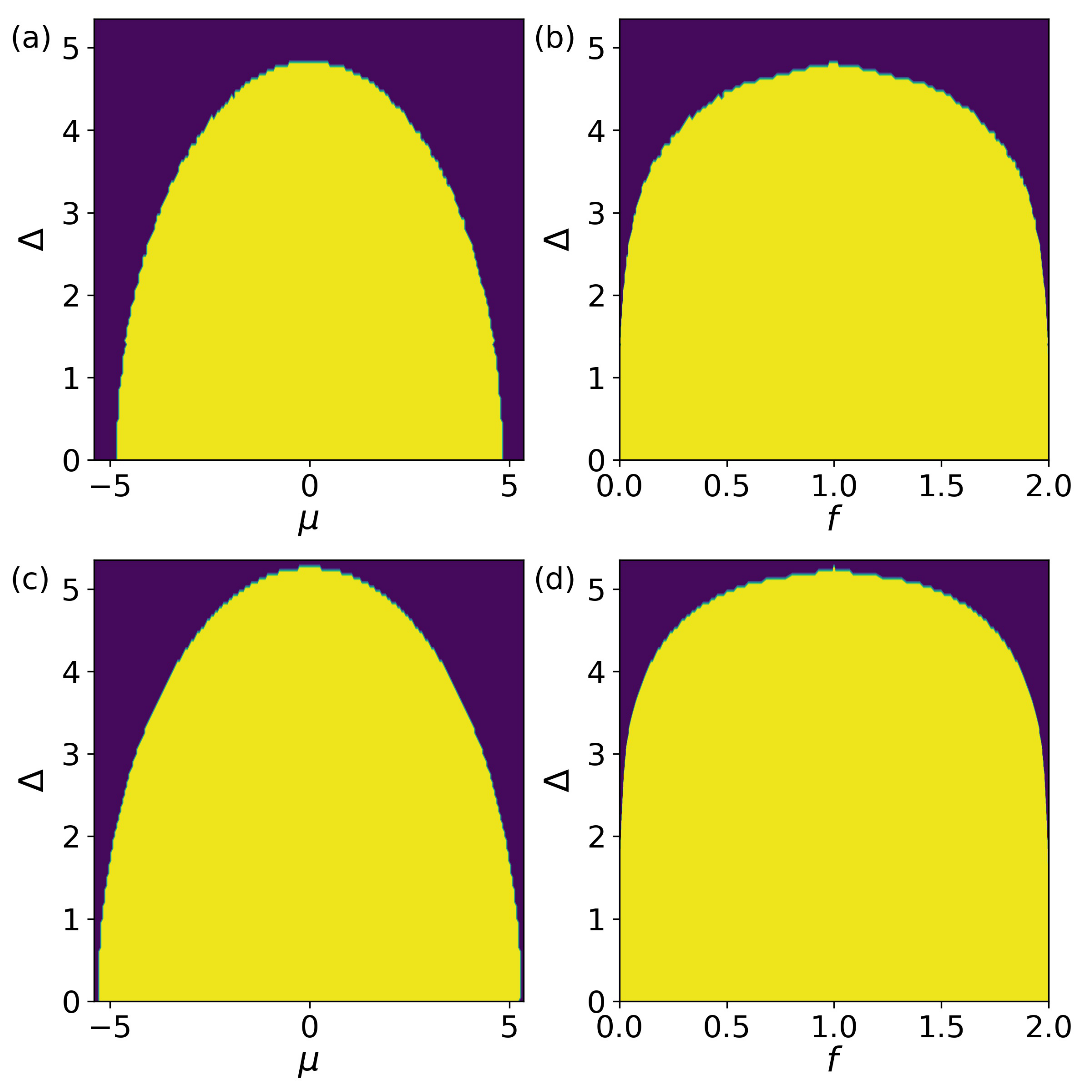}
\caption{(Color) Phase diagram of a 16-layer (100) supercell of a superconducting magnetic Weyl semimetal in (a) Weyl phase I and (c) Weyl phase II as a function of superconducting gap function $\Delta$, and chemical potential $\mu$. (b) and (d) are the same as (a) and (b) except verse filling factor $f$. The area of nontrivial phase in (c) and (d) is larger than that in (a) and (b).}
\label{fig2}
\end{center}
\end{figure}

To characterize the effect of the superconducting pairing potential on the electronic band structure, we perform a Wilson loop analysis on a $16\times 1 \times 1$ supercell with periodic boundary conditions to obtain the Chern number~\cite{Yu11}. By tuning the chemical potential $\mu$ and gap function $\Delta$, we map out the phase diagram, as shown  in Fig. \ref{fig2}, for the various Weyl phases. The purple and yellow regimes denote topological trivial and nontrivial phases, respectively. Since time-reversal symmetry is broken, this system belongs to the class D topological superconductors \cite{Schnyder08}.

Figure \ref{fig2}(a) and \ref{fig2}(b) present the phase diagram of Weyl phase I as a function of superconducting pairing strength verse the chemical potential and filling factor, respectively. Interestingly, the nontrivial topological phase spans fillings from $0$ to $2$ even for small $\Delta$. Demonstrating that even an infinitesimal small Cooper pairing potential facilitates a finite band inversion. Interestingly, in comparing these results to the phase diagram for Weyl phase II [Fig. \ref{fig2} (c) and (d)] we find the regime of nontrivial topological phase is larger in Weyl phase II than Weyl phase I. This key difference is what allows us to restrict a topologically nontrivial superconducting phase to a magnetic domain wall. That is, by choosing a $(\mu, \Delta)$ to be nontrivial for Weyl phase II, but trivial for Weyl phase I, we can design a domain wall structure where away from the domain the magnetization is aligned along the $z$-axis while at the domain wall the magnetization is pointing along the $x$ direction, thus inducing a local topologically nontrivial phase.

The phase boundary may also be obtained analytically from the Hamiltonian in momentum space
\begin{equation}\label{eq:ham_momentum}
H\!\!=\!\!\!\!\begin{bmatrix}
\!-\!\mu\!+\!m_z\!+\!h_z & h_x\!-\!ih_y & 0 & \Delta \\
h_x\!+\!ih_y & \!-\!\mu\!-\!m_z\!-\!h_z & \!-\!\Delta & 0 \\
0 & \!-\!\Delta^* & \mu\!-\!m_z\!+\!h_z & \!-\!h_x\!+\!ih_y \\
\Delta^* & 0 & \!-\!h_x\!-\!ih_y & \mu\!+\!m_z\!-\!h_z\! 
\end{bmatrix},
\end{equation}
by recognizing that the determinant is the product of the eigenvalues. Therefore, if the bands are gapless $\det(H)=0$ otherwise the system is gapped. Using this fact, the phase boundary can easily be determined as a function of $\mu$ and $\Delta$.

The determinant of the Hamiltonian in Eq.~(\ref{eq:ham_momentum}) is $D(h_x,h_y,h_z)=(h_x^2+h_y^2+h_z^2+m_z^2-\Delta^2-\mu^2)^2+4h_y^2\Delta^2+4h_z^2(\Delta^2-m_z^2)$. To analyze the local minima of $D$ as a function of $h_x$, $h_y$, and $h_z$ we first start by tuning $h_x$. The minimum of $D$ is found at $h_x^2=\Delta^2+\mu^2-m_z^2-(h_y^2+h_z^2)$ with minimum $D_{min}^x=4h_y^2\Delta^2+4h_z^2(\Delta^2-m_z^2)$. For Weyl phase II $[m_z=0]$, the minimum value is $0$ for $h_y=h_z=0$, which implies the bands are always gapless so long as the minimum condition can be reached. Therefore, the phase boundary corresponds to a critical point where $h_x^2+h_y^2+h_z^2=\Delta^2+\mu^2$. The maximum value of $h_x^2+h_y^2+h_z^2$ is $(2t_x+4t+m_x)^2$ located at a corner [$(\pi,\pi,\pi)$] of the Brillouin zone when $t'<t$, with a corresponding phase boundary $\Delta^2+\mu^2=(2t_x+4t+m_x)^2$. The analysis of Weyl phase I $[m_x=0]$ were split into two cases: $\Delta>m_z$ and $\Delta<m_z$. For $\Delta>m_z$, following our steps above the minima condition obeys $4h_z^2(\Delta^2-m_z^2)\ge0$, with the phase boundary given by $\Delta^2+\mu^2=(2t_x+4t)^2+m_z^2$. For $\Delta<m_z$, the phase boundary is slightly perturbed away from $(2t_x+4t)^2+m_z^2$. Comparing the phase boundary for Weyl phase I and II we find $(2t_x+4t+m)^2>(2t_x+4t)^2+m^2$ when taking $m_x=m_z=m$. Therefore, to produce a topologically nontrivial domain wall, parameters must be chosen to respect the inequality $(2t_x+4t+m)^2>\Delta^2+\mu^2>(2t_x+4t)^2+m^2$.

\begin{figure}[t]
\begin{center}
\includegraphics[clip = true, width =\columnwidth]{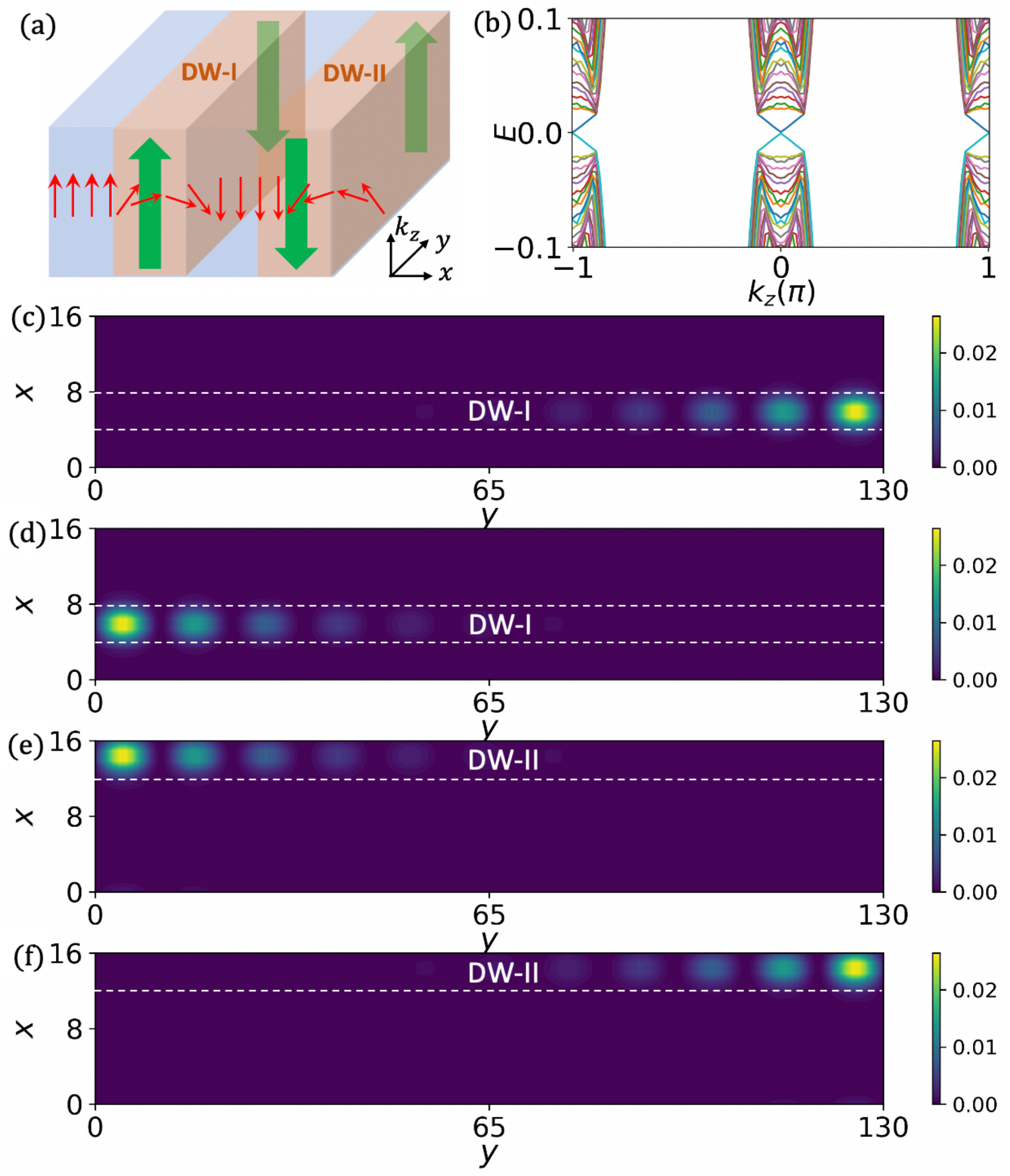}
\caption{(Color) (a) schematic of the two domain wall geometry with the magnetization (red arrows) varying along the $x$-axis. The green arrows indicate the propagation direction of the chiral Majorana edge modes. (b) Electronic band dispersions along $k_z$ in the Brillouin zone for a rectangular slab of superconducting magnetic Weyl semimetal. (c), (d), (e), and (f) show the amplitude of wavefunction at $k_z$ momentum $-0.926\pi$ and $0.074\pi$ for the pairs of linearly dispersing bands near the Fermi level. The location of the domain wall is marked by the white dash lines. 
}
\label{fig3}
\end{center}
\end{figure}

Figure~\ref{fig3}(a) shows a schematic of the domain wall geometry where the magnetization (red arrows) varies along the $x$ direction. The magnetization rotates clockwise by $\pi$ across the first domain wall (DW-I) and then rotates by $\pi$ again across the second domain wall  (DW-II) allowing for periodic boundary condition along $x$ and avoiding spurious defect states introduced by an edge. Since the magnetization is uniform along $y$ and translational invariance is only broken along the $x$-axis, the domain walls are two-dimensional, covering the $y$-$z$ plane. To study the edge states in the $y$-$z$ plane, we need to take an open boundary condition in the $y$ or $z$ dimension.
For the remaining discussion, let $t_x=0.4t$, $m=0.5t$, $\mu=5t$, and $\Delta=0.3t$ all of which satisfy the condition for localized Majorana modes. 

Figure \ref{fig3}(b) shows the electronic band structure along the $k_z$ direction for the slab geometry with $N_x=16$ and $N_y=130$. DW-I and II are constrained at $x\in(4,8)$ and $(12,16)$. Two pairs of linearly dispersing bands are clearly seen crossing the Fermi level at $0$ and $\pm \pi$ along $k_z$, indicating the existence of topological edge modes. For the first pair, the corresponding Bogoliubov wavefunction amplitude [Fig. \ref{fig3}(c) and \ref{fig3}(d)] is highly localized at the edges of the DW-I, decaying rapidly into the bulk. The propagation direction of each Majorana mode is further elucidated from the sign of the Fermi velocity and indicated by the green arrows in Fig. \ref{fig3}(a). Similarly, the amplitude of the wavefunction for the second pair [Fig. \ref{fig3}(e) and \ref{fig3}(f)] is located at the edges of DW-II, with propagation directions opposite to those pinned to DW-I. Fundamentally, this is driven by the fact that  DW-I and DW-II have opposite Chern numbers due to the opposing magnetizations on the domains. As a result the winding numbers in the Wilson loop analysis mutually cancel, as discussed further in the supplemental material \cite{supp}. Furthermore, if  a single domain wall with open boundary conditions is equivalently examined, a finite winding number is found~\cite{supp}.

\begin{figure}[t]
\begin{center}
\includegraphics[clip = true, width = \columnwidth]{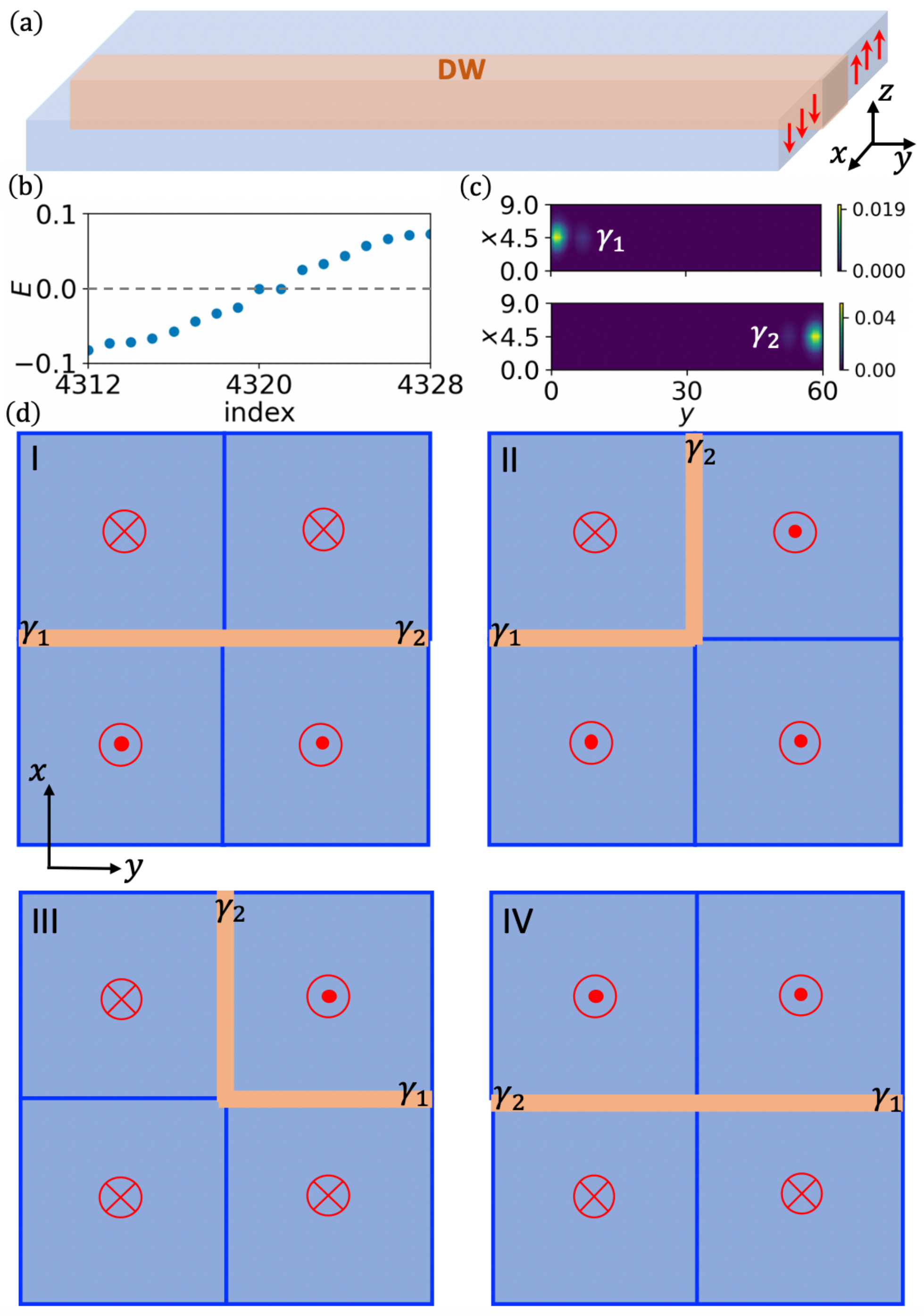}
\caption{(Color) (a) An atomically-thin film slab of a magnetic Weyl semimetal with a one dimensional magnetic domain wall. (b) The eigen-energy spectrum close to the Fermi level with two zero-energy states corresponding to localized Majorana bound states. (c) Projected wavefunction amplitude of the Majorana bound states in real space. (d) Proposed four step Majorana braiding protocol. Direction of the external magnetic field along the $z$-axis in the various quadrants of the magnetic Weyl semimetal slab is  indicated in red.}
\label{fig4}
\end{center}
\end{figure}

We now shift our discussion to the case of an atomically-thin film breaking $z$-axis translational invariance, as depicted in Figure \ref{fig4}(a), to elucidate our Majorana mode braiding protocol. In this case, there is a single one-dimensional domain wall and running along the $y$-axis.  Specifically, we utilize a $9\times60\times4$ super cell with $\mu=3.5t$ and $\Delta=0.9t$ and study its nontrivial topology directly by diagonalizing the Hamiltonian in the real space. Here, the magnetization is positive (negative) in the upper (lower) half plane centering the domain wall at $x=0$. Figure \ref{fig4}(b) shows the eigen-energy spectrum close to the Fermi level with two zero-energy states, suggesting the existence of two localized Majorana bound states. Projecting the wavefunction of these levels onto the second layer in the $z$ direction [Fig. \ref{fig4}(c)] we confirm the presence of the localized edge modes. For the projected wavefunction on the various layers in the system, see the appendixl~\cite{supp}.

From the two examples above, we clearly see that nontrivial topology comes about at the interface between two differing magnetization domains in the Weyl semimetal. Therefore, by patterning various magnetic domain topologies with external magnetic fields we can control and manipulate the presence and position of the localized Majorana modes. Thereby, providing a mechanism by which to facilitate braiding.

Figure \ref{fig4}(d) presents our proposed Majorana braiding protocol using a magentic Weyl semimetal thin-film. The surface of the film is divided into four quadrants with independently tunable magnetic fields. To initialize a state, the upper and lower half-planes are oppositely polarized creating a domain wall with Majorana bound states at either end, labeled as $\gamma_1$ and $\gamma_2$ (see Fig. \ref{fig4}(d) I). To exchange the states, the magnetization in the upper right quadrant is slowly reversed, moving the domain wall and the position of $\gamma_2$ to the top of the film (see Fig. \ref{fig4}(d) II). Next, the magnetization in the lower half-plane is slowly reversed, leading to the flipping of the $\gamma_1$ position from the $-\hat{x}$ edge to $+\hat{x}$ edge of the sample (see Fig. \ref{fig4}(d) III). During this process the position of $\gamma_2$ does not change. Finally, the magnetization of the upper left quadrant is inverted, recovering the initial domain wall, but with the position of $\gamma_1$ and $\gamma_2$ swapped, completing one braiding operation (see Fig. \ref{fig4}(d) IV).

Due to the numerous experimental demonstrations inducing and controlling magnetic domain walls in real quantum materials~\cite{RevModPhys12,Schellekens12,Koyama12,PRL21}, we are optimistic that the same level of robustness and precision can be achieved in the magnetic Weyl semimetals. Furthermore, a recent work by Xu {\it et al.}~\cite{Xu21} on $\mathrm{CeAlSi}$ has exemplified this claim. The key ingredients required by our proposal, such as strong anisotropy and fine tuning of the chemical potential, have been recently experimentally suggested in a several magnetic Weyl semimetals [including $\mathrm{Co_3Sn_2S_2}, \mathrm{Mn_3Sn}, \mathrm{Co_2MnGe}$ and $\mathrm{CeAlSi}$] or is readily realized by chemical doping  or  electrostatic gating thin film samples. 

Lastly, the presence of superconductivity is essential for the realization of Majorana modes on the domain walls. The superconductivity and magnetic Weyl phase can coexist intrinsically in materials, termed Weyl superconductors~\cite{Meng12}. Presently, there is mounting evidence suggesting $\mathrm{UTe_2}$ and $\mathrm{MoTe_2}$ to be Weyl superconductors~\cite{Ran684,Hayes20,Wang534}. For generic magnetic Weyl semimetals, superconductivity can be induced by proximity effect, through heterostructuring with other bulk superconductors, such as $\mathrm{Nb}$. 

The four ingredients of this topological superconducting domain wall are anisotropic band structure, band topology, magnetism and superconductivity. In principle, the materi- als option does not restrict to the magnetic Weyl semimetal. However, these ingredients except for superconductivity is generically available in this material which makes it a natu- ral platform to search for such domain walls.

{\it Conclusion} -- We have demonstrated the existence of the chiral Majorana edge modes or zero-energy Majorana bound states on the domain walls of an anisotropic magnetic Weyl semimetal with intrinsic superconductivity or using a superconducting proximity effect. Because the magentic domain wall emerges at the interface between two differing magnetic domains, they are easily controlled and manipulated by external magnetic fields. Finally, we have proposed a straightforward braiding protocol of zero-energy Majorana bound states.

\begin{acknowledgments}
{\it Acknowledgments} -- We thank Benedikt Fauseweh and Sarah Grefe for useful discussions. This   work   was   carried   out   under   the   auspices of  the  U.S.  Department  of  Energy  (DOE)  National Nuclear  Security  Administration  under  Contract  No. 89233218CNA000001. 
 The Weyl semimetal part of the work was supported by the U.S. DOE BES EFRC ``Center for the Advancement of Topological Semimetals'' Project. 
 The topological superconductivity part of the work was supported by 
UC Laboratory Fees Research Program (Grant Number: LFR-20-653926).
 C.L. was supported by LANL LDRD Program. It was also supported in part by the Center for Integrated Nanotechnologies, a DOE BES user facility, in partnership  with  the  LANL  Institutional  Computing  Program for computational resources.
\end{acknowledgments}

\begin{widetext}

\section{Single domain wall}
The effective Hamiltonian of a magnetic Weyl semimetal including the mean-field pair potential for a three-dimensional lattice  is given by
\begin{eqnarray}\label{H0}
H_0 &&=\sum_j -\mu c_j^{\dagger}c_j+\left[ -t_xc_j^\dagger \sigma_x c_{j+\hat{x}}-t(c_j\sigma_x c_{j+\hat{y}}+c_j\sigma_x c_{j+\hat{z}})-it'( c_j^\dagger \sigma_y c_{j+\hat{y}}+c_j^\dagger \sigma_z c_{j+\hat{z}})+H.c.\right] \nonumber \\
&&+m_xc_j^\dagger \sigma_x c_j+m_zc_j\sigma_z c_j+\Delta c_{j\uparrow}^\dagger c_{j\downarrow}^\dagger+\Delta^* c_{j\downarrow}c_{j\uparrow}, 
\end{eqnarray}
where $\mu$ is the chemical potential, $t_x$, $t$ and $t'$ are the spin-orbit coupling coefficients, $m_x$ and $m_z$ are the projected Zeeman field strengths along the $x$  and $z$ axes.  

\begin{figure}[h]
\begin{center}
\includegraphics[clip = true, width =0.5\columnwidth]{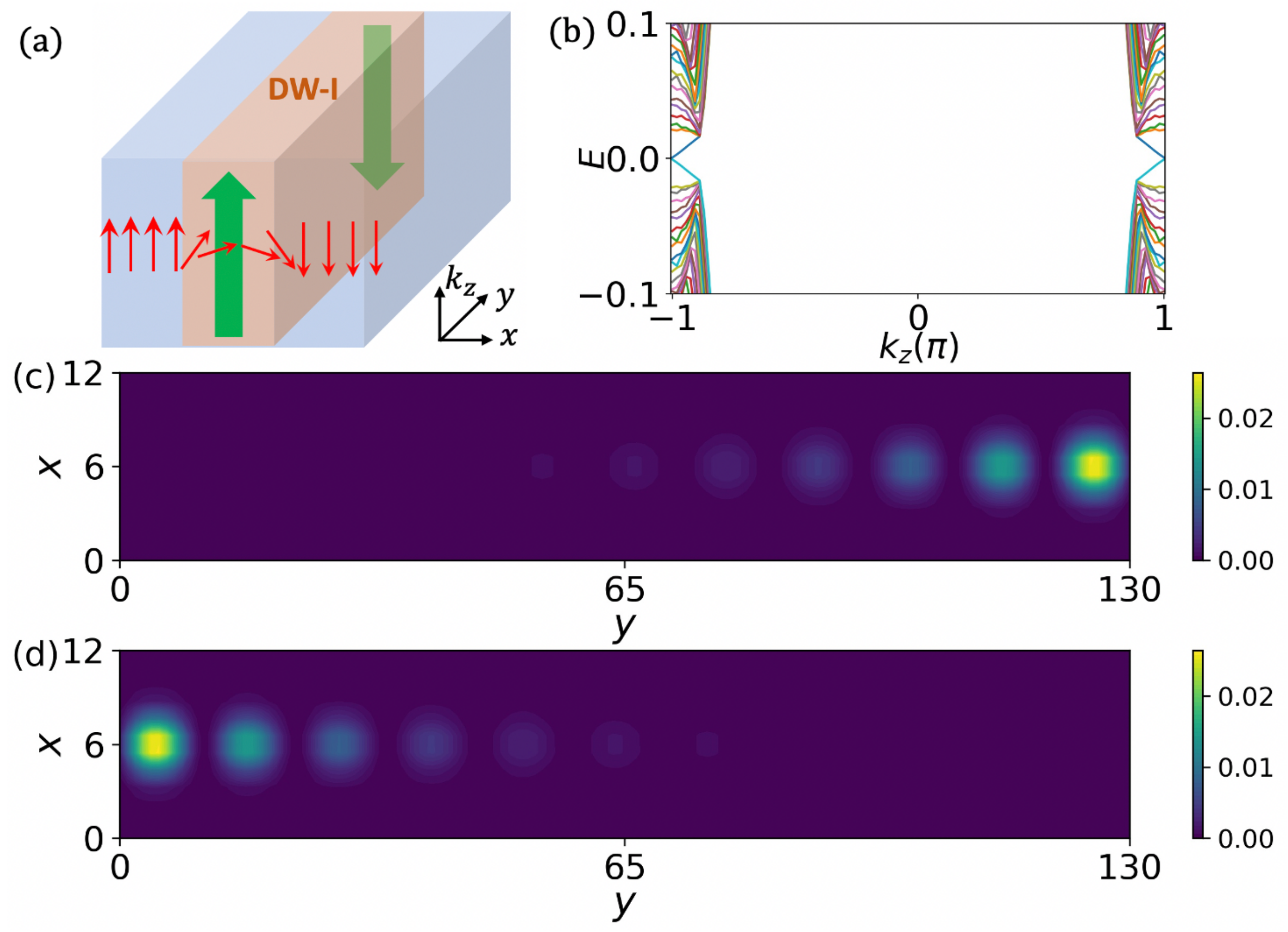}
\caption{(color online) (a) Schematic of a single one-dimensional domain wall with the magnetization (red arrows) varying along the $x$-axis. The green arrows indicate the propagation direction of the chiral Majorana edge modes. (b) Electronic band dispersions along $k_z$ in the Brillouin zone for a rectangular slab of superconducting magnetic Weyl semimetal. (c) and (d) show the amplitude of wavefunction at $k_z$ momentum $-0.926\pi$ for the pair of linearly dispersing bands near the Fermi level.}
\label{figs1}
\end{center}
\end{figure}

Figure~\ref{figs1}(a) shows a typical single domain wall within a $12\times130\times1$ supercell with the magnetization (red arrows) rotating along the $x$-axis such that there is a $\pi$ clockwise rotation across the the domain wall (DM-I). In contrast to Fig.~$3$ of the main text, open boundary conditions in the $x$ direction is applied, dividing the system into three segments (each having 4 layers along $x$-direction). Here, we used the same parameters, $t_x=0.4t, t'=0.4t, m=0.5t, \mu=5t, \Delta=0.3t$ and $N_y=130$, as the main text. 

Figure~\ref{figs1}(b) presents the electronic band structure along the $k_z$ direction in the Brillouin zone. Two pairs of linearly dispersing bands are clearly seen crossing the Fermi level at $0$ and $\pm \pi$ along $k_z$, indicating the existence of topological edge modes. For the first pair, the corresponding Bogoliubov wavefunction amplitude [Fig.~\ref{figs1}(c) and \ref{figs1}(d)] is highly localized at the edges of the DM-I, decaying rapidly into the bulk. The propagation direction of each Majorana mode is further elucidated from the sign of the Ferml velocity and and indicated by the green arrows in Fig.~\ref{figs1}(a).

\section{Wannier Center Curves}
Figure~\ref{figs2} compares the evolution of the Wannier function center with $k_y$ momentum for the cases of a single domain wall [Fig.~\ref{figs1}(a)] and two domain walls [Fig. 3 of the main text]. For the single domain wall case, the Chern number is $C=1$ since the Wannier center traverses from $-\pi$ to $\pi$ once, indicating the existence of a single chiral edge mode~\cite{Yu2011}. For the two domain case, the Chern number is zero, since one winding goes through $2\pi$ and the other goes to $-2\pi$, thus mutually canceling. These two opposing winding numbers correspond to the opposite chirality of the chiral Majorana edge mode of the domain walls shown in Fig.~$3$ of the main text.  

\begin{figure}
\begin{center}
\includegraphics[clip = true, width =0.7\columnwidth]{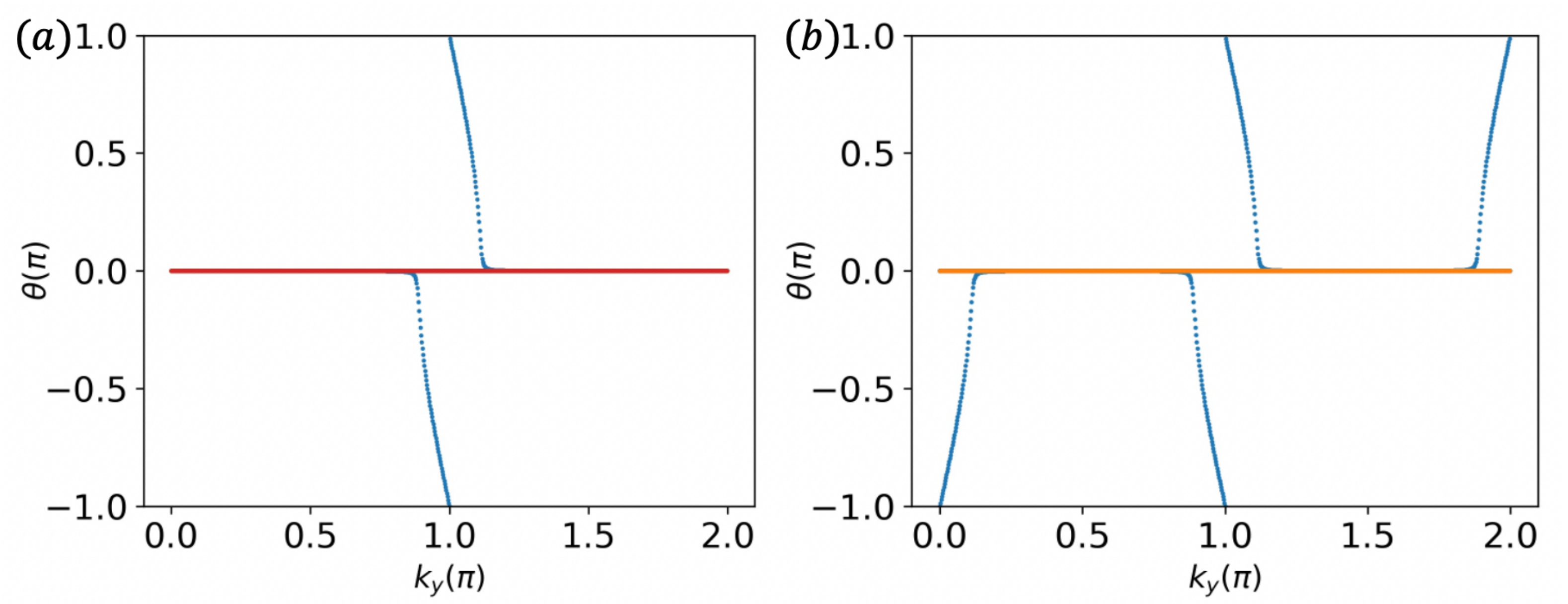}
\caption{(color online) The evolution  of the Wannier center curves along the $k_y$ axis for a  (a) single domain wall and (b) two domain walls.}
\label{figs2}
\end{center}
\end{figure}

\section{Atomically-thin Film of Magnetic Weyl Semimetal}
To elucidate Majorana mode generation we consider to cases: In. the first case,  the magnetization is equal and opposite in the upper and lower half-planes of a $9\times60\times4$ thin-film super cell of a typical magnetic Weyl semimetal. In the second case,  the magnetization is uniform across the whole plane. For both cases we study its electronic structure and possible nontrivial topology directly by diagonalizing the Hamiltonian in the real space using open boundary conditions and the following parameters $t_x=0.4t$, $t'=0.4t$, $m=0.5t$, $\mu=3.5t$, and $\Delta=0.9t$. In the first case, a single one-dimensional domain wall is formed running along the $y$-axis centered at $x=0$ [Fig.~\ref{figs3}(a)] with the magnetization rotating by $\pi$ across the domain wall. As a consequence, the corresponding electronic structure exhibits two zero-energy levels [Fig.~\ref{figs3}(b)]. In contrast, for a uniform magnetization, no such modes are present [Fig.~\ref{figs3}(c)].

\begin{figure}[h]
\begin{center}
\includegraphics[clip = true, width =0.7\columnwidth]{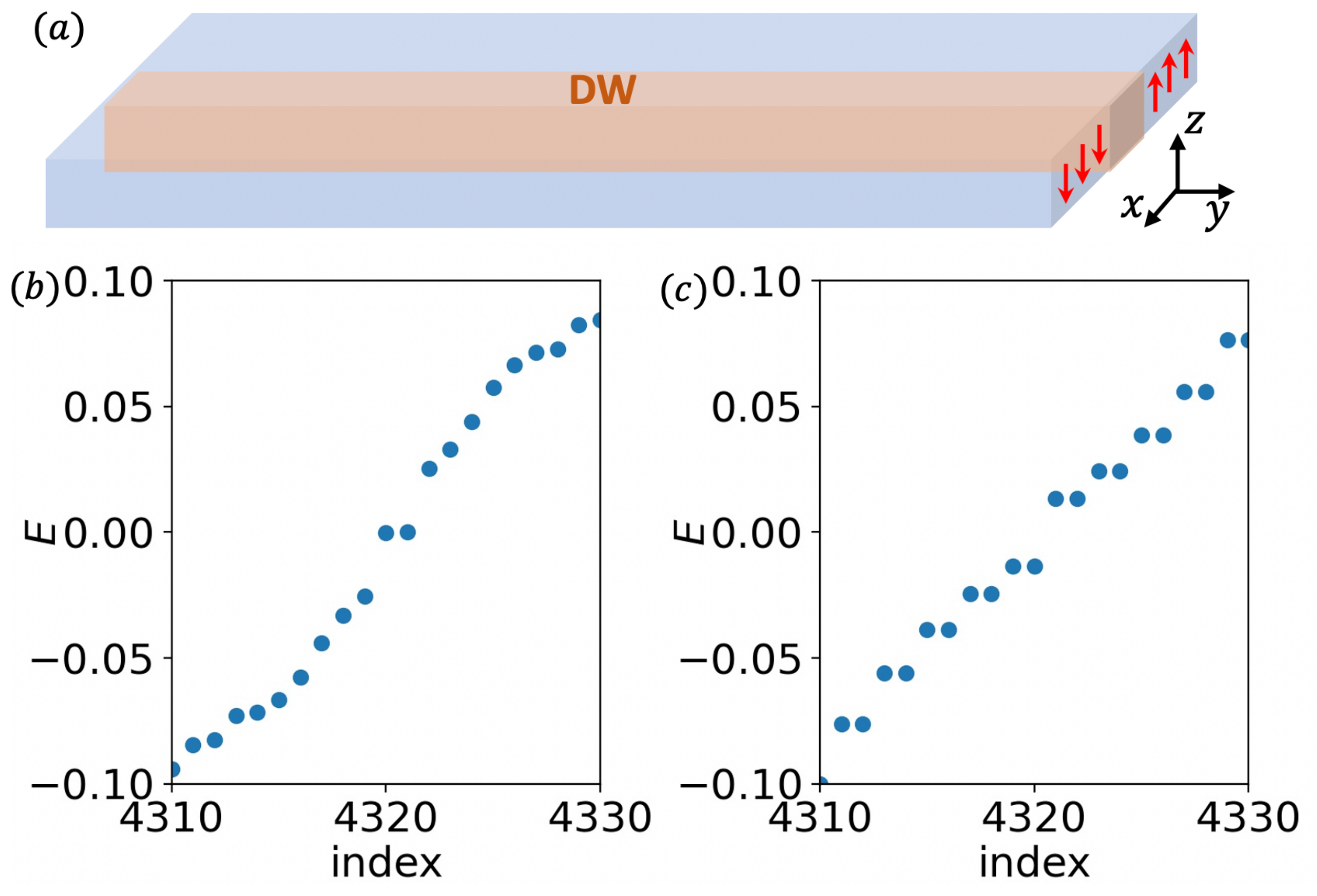}
\caption{(color online) (a) An atomically-thin film slab of a magnetic Weyl semimetal with a one-dimensional magnetic domain wall. The eigen-energy spectrum close to the Fermi level with (b) and without (c) the domain walls. The panel (b) shows clearly the existence of two zero-energy modes corresponding to Majorana bound states}
\label{figs3}
\end{center}
\end{figure}

Figure \ref{figs4} presents the corresponding real space wavefunction amplitudes of the zero-energy modes. To capture an individual localized Majorana mode at the edge of the sample, we apply a unitary transformation to resolve the weak coupling between the edge states from the finite size effects, that is,  $\psi_1=(\psi_{+\epsilon}+\psi_{-\epsilon})/\sqrt{2}$ and $\psi_2=(\psi_{+\epsilon}-\psi_{-\epsilon})/\sqrt{2}$, where $\psi_{\pm\epsilon}$ is the spatially dependent Bogoliubov quasiparticle wavefunction $(u_\uparrow, u_\downarrow, v_\uparrow, v_\downarrow)$ for the level $\pm\epsilon$. Figure \ref{figs4}(a) - (d) and Fig. \ref{figs4}(e) - (h) show the projection of $|\psi_1|^2$ and $|\psi_2|^2$ on the four layers along the $z$-axis. Clearly, the Majorana modes are localized at either end of the domain wall.

\begin{figure}
\begin{center}
\includegraphics[clip = true, width =0.8\columnwidth]{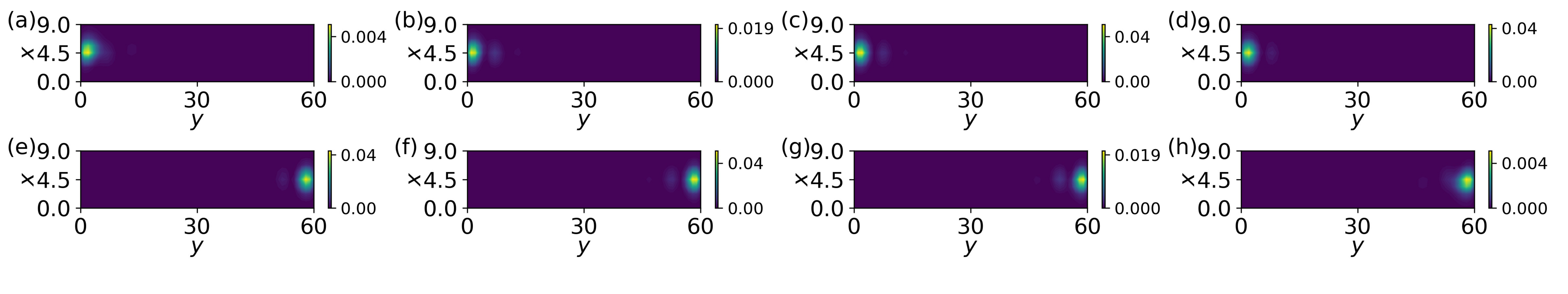}
\caption{(color online) Projected wavefunction amplitude of the left (a) - (d) and right (e) - (h) Majorana bound states for the various layers of the thin film in real space.}
\label{figs4}
\end{center}
\end{figure}

\end{widetext}

\bibliography{references}

\end{document}